\documentclass[a4paper, 12pt]{article}

\usepackage[utf8]{inputenc}
\usepackage[english]{babel}
\usepackage{amssymb, amsmath}

\usepackage{graphicx}
\graphicspath{{img/}}
\DeclareGraphicsExtensions{.pdf,.png,.jpg}

\begin{document}

\title{Machine Learning in Problems of Automation of Ultrasound Diagnostics of Railway Tracks}
\author{Andrey Igonin \\ Ulyanovsk State Technical University \\ a.igonin@ritg.ru \and Vitaliy Ulybin \\ Russian IT Group \\ v.ulybin@ritg.ru}
\date{}
\maketitle

\abstract{The article presents the system architecture for automatic decoding of raiway track defectograms in real time. The system includes an ultrasound data processing module, a set of neural networks classifiers, a decision block. Preprocessing of data includes affine transformations of measurement information into a format suitable for the operation of a neural network, as well as a combination of information on measurement channels, depending on the type of defect being defined. The classifier in built on a convolutional neural network. The proposed solution can be effectively implemented on a modern elemental  basis for performing parallel computing, including TPUs and GPUs.}

\section{Introduction}

The increase in speed and intensity of movement of trains on the tracks, caused an increase in quality control requirements. At the same time, the time of the control, data processing, analysis and decision-making on the admission of the railway is reduced. Integrity monitoring includes the detection of any internal defects: cracks, voids, delamination, loss of integrity of connections, etc. It is crucial to prevent disasters and encure the stability of rail transport.

In this regard, the development of automation systems, new models and methods for finding defects in real-time mode is relevant. One of the classic means of scanning railway rails is the use of ultrasound. Various methods are used to decipher ultrasonic defectograms of the rails. Starting from manual, methods based on the analysis of physical principles and ending with tools for analyzing various types of graphical presentation. This article discusses a method for analyzing and identifying defects using machine learning tools. A graphical representation of ultrasonic defectograms is considered as data for analysis.

\section{Subject matter and system requirements}

During the last 5 years, ultrasonic devices have appeared on the market that allow for high-speed testing of railway rails. Controls systems include the detection and registration of defects in real time, as well as the complete recording of scanned data. These data can be used for periodic monitoring of the state of the railway, by identifying defects in the controlled  object with the necessary accuracy and reliability.

With the manual method of decoding the visual appearance of defectograms obtained during ultrasound testing, experts are involved. In this case, the main disadvantages are the presence of factors that influence the results obtained. These factors can be: qualification, affentiveness, physical and mental state of a person, etc. Given the large volume and complexity of the visual presentation of data, manual processing performance is very low.

Moreover, in recent years, to facilitate the work of specialists, tools have been developed to facilitate the proccessing of large amounts of data \cite{Armbruster}, such as:

- data output in a convenient view;

- filtering noise signals, for a clearer visualization of areas that may contain defects;

- means of classification based on harsh conditions.

All these tools accelerate data processing, but in modern conditions do not meet the requirements:

- by definition time;

- by the percentage of defects detected;

- at the time of the decision;

- in automated systems, the presence of stringent conditions does not allow to determine all types of defects in the presence of various characteristics of detects;

- there is no possibility to conduct an analysis in the mode / speed of receipt of data from the flaw detector sensors.

These funds also do not allow for the full and for all identified and possible cases of comparative and periodic monitoring.

The level of development of machine learning systems, in particular, neural network algorithms and hardware, make it possible to organize their use for solving the above problem. Development of an automated identification systems (AIS) based on a neural network, which will allow to solve the problem of identifying and classifying defects without imposing restrictions on the input data \cite{Bettayeb2008}. Such a system allows to increase the effeciency of rail monitoring, to organize control in the mode of data entry into the systems, as well as to increase the reliability and accuracy of the results obtained by minimizing the influence of subjective factors.

To achieve its goals, AIS must meet the following requirements:

- work with single-channel and multichannel ultrasound control systems;

- the ability to identify and classify all types of defects in the subject area;

- the computational complexity of the algorithms used should ensure that results are obtained in real time when implemented on modern hardware platforms (including parallel ones) with the speed of movement of scanning platforms up to 110 km/h.

\section{System description}

Scanning probes are located on a special mechanical skid on the front wheel pair (Fig.1)

\begin{figure}[h]
	\includegraphics[width=\linewidth]{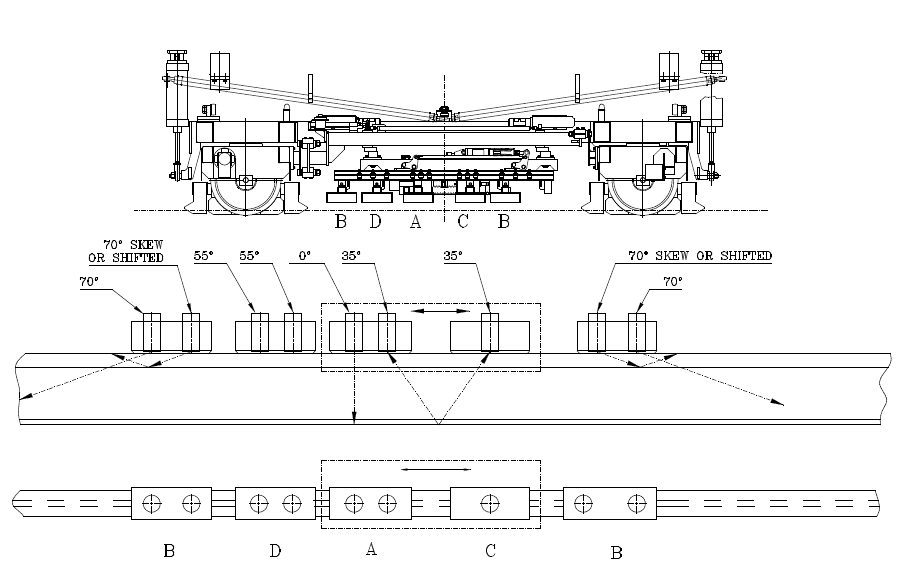}
	\caption{Ultrasound diagnosis system}
\end{figure}

When determining the location of the defect in the calculations, the position of each probe relative to the zero position of the system is taken. In this case, the angle of inclination of the sensor is not taken into account, therefore the indication of the depth of the defect and the real position along the scanning axes do not correspond to its actual location in the rail being diagnosed. This is done to simplify the identification of defects, as show in Fig.2.

\begin{figure}[h!]
	\includegraphics[width=\textwidth]{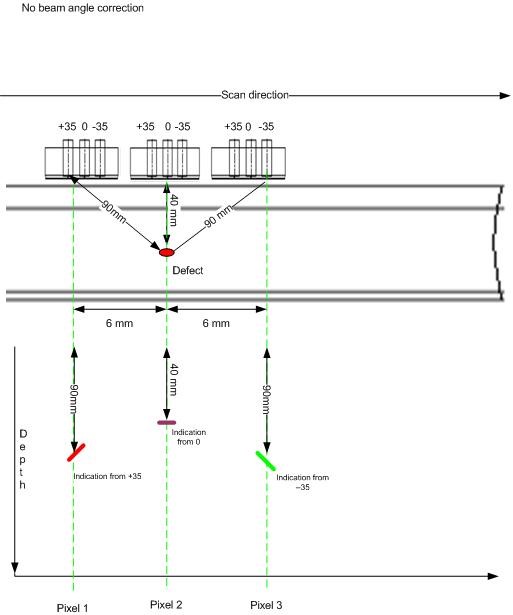}
	\caption{Identification of defects}
\end{figure}

The indications exicting in the railway can be divided into two large classes: technological and defects. To technological include: bolted joints, joints and welds. The defects are: horizontal crack of the rail head, delamination of the rail head, detachment of the rail bottom, crack in the neck of the rail, emptiness in the weld, vertical crack of the rails, cracks in bolted joints (star  cracks). Also, defects can be displaced not only in a vertical plane, but also have a horizontal displacement relative to the central axis in the rail head.

Each type of defect gives an indication not on all channels of the probes, but only at certain angles (Table 1). Using this feature allows you to combine information on the measuring channel in accordance with the type of defect, this can significantly reduce the amount of information supplied to the appropriate classifier and thereby significantly increase the overall speed of the system.

\begin{table}[h]
	\caption{Combining sensor data}
	\begin{tabular}{| c | c | c |}
		\hline
		N & Sensor combination & Detectable defect types \\
		\hline
		1 & $0^o$ & Horizontal cracks in the head, neck and foot rail \\
		\hline
		2 & $\pm70^o$ & Transverse cracks in the head, neck and foot rail \\
		\hline
		3 & $0^o, \pm35^o, \pm70^o$ & Bolt holes \\
		\hline
		4 & $0^o, \pm35^o$ & Inclined cracks \\
		\hline
		5 & $\pm55^o$ & Cracks in the head, neck and foot rail \\
		\hline
	\end{tabular}	
\end{table}

The proposed general scheme of the process of decoding a defectogram is shown in Fig.3. Its main stages are the stage of preliminary processing of ultrasound data, including sampling data from the sensors of the ultrasound system, bringing the amplitude values into a range and combining the data, as well as the stage of searching for defects and making a decision.

\begin{figure}[h!]
	\includegraphics[width=\textwidth]{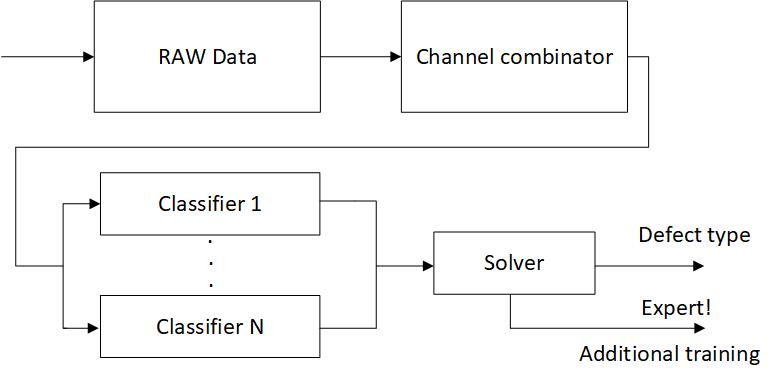}
	\caption{Architecture of the automatic recognition system of railway bed defects}
\end{figure}

The stack of measurement information is used to select the source data. The data of each channel are recorded sequentially in the tensors of the same size.

In the next step, the data of individual channels that are included in the sample are combined into a single structure in accordance with Table 1, which can be used to search for signs of defects. Defects are searched for in parallel by specialized classifiers.

Decisions of specialized classifiers are transmitted to the input of the decisionmaking module, which assesses the probability of correctness of decision making by the composition of classifiers \cite{Yaman2008}, and in case of low level of confidence intervals, delegates this decision to an expert. The solutions specified by the expert are further used for the further training of neural networks of the classifier, thereby gradually improving the quality of the entire system.

To solve the problem associated with the detection and classification of defects as an object of technical diagnostics, convolutional neural networks (CNN) were used, which allowed:

1. Effectively parallelize computational processes, which made it possible to organize the work of the system in real time.

2. Implement a mechanism for further training classification.

3. To ensure high computational performance of the neural network due to the use of a much smaller number of parameters compared to other neural network architectures.

4. To provide resistance to turns and shifts of recognizable frames.

\section{Conclusion}

The development of systems that provide decoding of ultrasonic defectograms of railway rails in automatic mode at high scanning speeds is a difficult task. At the same time, the models used to solve this problem at present have a number of limitations that do not allow solving it completely.

The application of the proposed system of models in the development of an automated system for decoding ultrasonic defectograms will improve the efficiency of the rail monitoring process, the reliability of the results obtained and the safety of rail transport.

\end{document}